\documentclass{IEEE_lsens}
\usepackage{textcomp}

\usepackage[noadjust]{cite}

\ifCLASSINFOpdf
\else
\fi
\usepackage[T1]{fontenc} 
\usepackage{amsmath}
\usepackage[cmintegrals]{newtxmath}
\usepackage{bm}

\usepackage{array}
\usepackage{url}
\ifCLASSINFOpdf
\else
\fi
\providecommand{\hypersetup}[1]{\relax}

\hypersetup{pdftitle={Bare Demo of IEEE\_lsens.cls for IEEE Sensors Letters},
pdfsubject={Typesetting},
pdfauthor={Michael D. Shell},
pdfkeywords={Class, IEEE, IEEE\_lsens, IEEE Sensors Letters, LaTeX, Typesetting, TeX}}
\usepackage[latin1]{inputenc}  
\usepackage[T1]{fontenc}
\usepackage{graphicx}          
\usepackage{psfrag}           
\usepackage{amsfonts}
\usepackage{amsmath,amssymb}
\usepackage{supertabular}
\usepackage{tikz}
\usepackage{url}
\usepackage{cite}

\usepackage[latin1]{inputenc} 

\usepackage{graphicx}
\usepackage{epstopdf}
\usepackage[process=auto]{pstool} 
\usepackage{psfrag}

\usepackage{bm}
\usepackage{bbm}
\usepackage{cite}
\usepackage{subfigure}

\begin{document}


\IEEELSENSarticlesubject{}

%
\title{Zero-Velocity Detection -- A Bayesian Approach to Adaptive Thresholding}

%
\author{\IEEEauthorblockN{Johan~Wahlstr\"om\IEEEauthorrefmark{1},~Isaac~Skog\IEEEauthorrefmark{2}\IEEEauthorieeemembermark{1},~Fredrik~Gustafsson\IEEEauthorrefmark{3}\IEEEauthorieeemembermark{2},~Andrew~Markham\IEEEauthorrefmark{1},~and~Niki~Trigoni\IEEEauthorrefmark{1}}
\IEEEauthorblockA{\IEEEauthorrefmark{1}Department of Computer Science, University of Oxford, Oxford OX1 2JD, UK\\
\IEEEauthorrefmark{2}S3 Research AB, 116 29 Stockholm, Sweden
\\
\IEEEauthorrefmark{3}Department of Electrical Engineering, Link\"oping University, 581 83 Link\"oping, Sweden\\
\IEEEauthorieeemembermark{1}Senior Member, IEEE \\
\IEEEauthorieeemembermark{2}Fellow, IEEE
}%
\thanks{Corresponding author: Johan Wahlstr\"om (e-mail: johan.wahlstrom@cs.ox.ac.uk).}
}
%
%
%


\IEEEtitleabstractindextext{%
\begin{abstract}
A Bayesian zero-velocity detector for foot-mounted inertial navigation systems is presented. The detector extends existing zero-velocity detectors based on the likelihood-ratio test, and allows, possibly time-dependent, prior information about the two hypotheses -- the sensors being stationary or in motion -- to be incorporated into the test. It is also possible to incorporate information about the cost of a missed detection or a false alarm. Specifically, we consider an hypothesis prior based on the velocity estimates provided by the navigation system and an exponential model for how the cost of a missed detection increases with the time since the last zero-velocity update. Thereby, we obtain a detection threshold that adapts to the motion characteristics of the user. Thus, the proposed detection framework efficiently solves one of the key challenges in current zero-velocity-aided inertial navigation systems: the tuning of the zero-velocity detection threshold. A performance evaluation on data with normal and fast gait demonstrates that the proposed detection framework outperforms any detector that chooses two separate fixed thresholds for the two gait speeds. 
\end{abstract}

\begin{IEEEkeywords}
Zero-velocity updates, foot-mounted inertial navigation, indoor localization, posterior odds ratio, adaptive thresholding.
\end{IEEEkeywords}}




\maketitle

\section{Introduction}

\begin{figure}[t!]
	\centering
	\includegraphics[width=0.9\columnwidth]{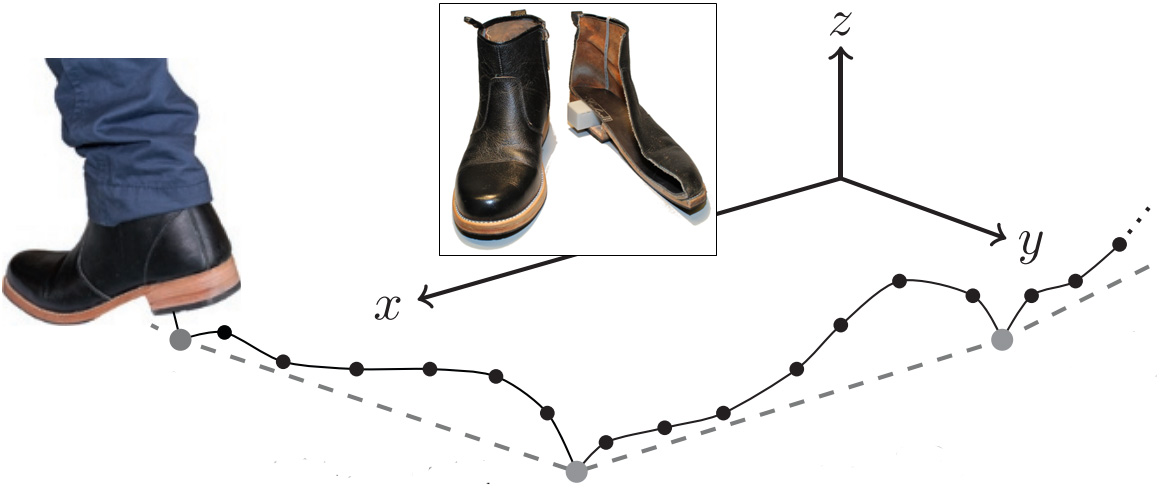}
	\caption{Illustration of the concept of zero-velocity-aided inertial navigation. Using foot-mounted inertial sensors, the motion of the foot is tracked. To mitigate the error growth, zero-velocity updates are used to correct the navigation solution. One of the remaining challenges is how to reliably detect the zero-velocity instances during different usage and gait conditions.}\label{F:step}
\end{figure}

Zero-velocity-aided inertial navigation is one of the most promising technologies for indoor positioning in environments without pre-installed infrastructure \cite{Norrdine2016,Muhammad2018,Wahlstrom2019}. The basic concept is illustrated in Fig. \ref{F:step}. A key component of this technology is the zero-velocity detector, which identifies when the sensors are stationary and zero-velocity updates can be applied. Hence, a variety of detectors have been proposed, many of which can be derived as generalized likelihood ratio tests \cite{Skog2010}.

One of the problems with existing zero-velocity detectors is that the optimal detection threshold varies significantly with the pedestrian's walking speed and motion mode (walking, jogging, running, etc.), the placement of the sensor, the type of shoe, and the walking surface. If the threshold on the likelihood ratio is too large, the detector will not be able to detect stationary instances when the user is running. If the threshold is too small, the detector will produce false zero-velocity instances \cite{Nilsson2012}. Consequently, several methods for designing adaptive thresholds have been proposed. The conducted studies have mainly focused on gait speed variations.

The most common approach to adaptive thresholding is to first use some heuristic or ad-hoc solution for estimating or classifying the speed or motion mode of the user. Based on the result, the detector selects a threshold value that has been optimized, using ground truth data, for that specific speed or motion class \cite{Tian2016,Park2016,Zhang2017,Ma2017,Wagstaff2017}. However, other methods for robust zero-velocity detection under varying gait conditions have also been explored. The authors in \cite{Li2012} used accelerometer measurements to detect the beginning and end of individual steps, and then applied different detectors and thresholds to different parts of the gait cycle; \cite{Wagstaff2018} made the detection using a long short-term memory (LSTM) neural network; \cite{Sun2018} designed zero-velocity detectors that used a hidden Markov model (HMM) to represent the different stages of the gait cycle; \cite{Liu2014} allowed the threshold to vary with the temporal variance of the accelerometer measurements; and \cite{Wang2018} held the threshold fixed while dynamically adapting the window length of the samples used to compute the detection statistic.

\IEEEpubidadjcol

Unfortunately, it is not clear how these methods relate to the existing theory on zero-velocity detection using likelihood ratio tests. Moreover, the proposed methods tend to require large data sets collected at different gait speeds and motion modes to calibrate the threshold values or other design parameters. Additionally, methods that attempt to infer speed or motion mode are always limited by the accuracy of the speed estimation or motion mode classification that precede the threshold selection. 

This paper presents a zero-velocity detector based on the posterior odds ratio. The proposed detection framework demonstrates how the threshold, used in established detectors based on the likelihood ratio, can be factorized as a product of (i) the inverse prior odds ratio, quantifying the prior probability of a zero-velocity detection; and (ii) a loss factor, quantifying the cost of incorrect detections. The primary contributions are:

\begin{itemize}
	\item A theoretical justification of adaptive zero-velocity detection within the established framework for zero-velocity detection based on the likelihood ratio test.
	\item An application-specific interpretation of the threshold used in traditional methods for zero-velocity detection.
	\item A discussion of how to model the hypothesis prior and the loss factor that specify the detection threshold.
\end{itemize}
The performance of the proposed detector is illustrated using a large data set collected at varying gait speeds\footnote{Reproducible  research:  The  data and the code used in the experiments are available under an open-source licence at www.openshoe.org.}.

\section{Bayesian Zero-Velocity Detection}

Consider the problem of determining whether an inertial measurement unit (IMU) is stationary or not given the measurements $\mathbf{z}_n\overset{_\Delta}{=}\{\mathbf{y}_k\}_{k=n}^{n+N-1}$, collected over $N$ sampling instances. Here, $\mathbf{y}_k$ denotes the inertial measurements at sampling instance $k$. The problem can be formalized as the binary classification problem of choosing between the two hypotheses
\begin{align}
	\begin{split}
		&{\cal H}_0:\text{ IMU is moving} \\
		&{\cal H}_1:\text{ IMU is stationary}.
	\end{split}
\end{align}
In what follows, a method for performing the classification by minimizing the conditional risk is described.

\subsection{Minimum-Error-Rate Classification}

The performance of a Bayesian detector is quantified using the conditional risk \cite{Duda2000}
\begin{equation}
R({\cal H}_i|\mathbf{z}_n)\overset{_\Delta}{=}\textstyle\sum_{j=0}^{1}\lambda_{ij}p({\cal H}_j|\mathbf{z}_n).
\end{equation}
The conditional risk $R({\cal H}_i|\mathbf{z}_n)$ is the expected incurred loss when deciding on hypothesis ${\cal H}_i$ given the data $\mathbf{z}_n$. Here, $\lambda_{ij}$ represents the loss incurred as a result of deciding on hypothesis ${\cal H}_i$ when the true hypothesis is ${\cal H}_j$. To minimize the conditional risk, we should decide on hypothesis ${\cal H}_1$ if and only if 
\begin{equation}
	\label{eq_posterioroddstest}
	\frac{p({\cal H}_1|\mathbf{z}_n)}{p({\cal H}_0|\mathbf{z}_n)}>\eta
\end{equation}
where $\eta\overset{_\Delta}{=}(\lambda_{10}-\lambda_{00})/(\lambda_{01}-\lambda_{11})$. Throughout the paper, all denominators will be assumed to be nonzero.

\subsection{Relation to Established Zero-Velocity Detectors}

By using $p({\cal H}_0)=1-p({\cal H}_1)$ and the factorization 
\begin{equation}
	\underbrace{\frac{p({\cal H}_1|\mathbf{z}_n)}{p({\cal H}_0|\mathbf{z}_n)}}_{\hspace*{-4mm}\text{Posterior odds ratio}}\!=\underbrace{\frac{p(\mathbf{z}_n|{\cal H}_1)}{p(\mathbf{z}_n|{\cal H}_0)}}_{\hspace*{-1mm}\text{Likelihood ratio}}\hspace*{1mm}\cdot\hspace*{-3mm}\underbrace{\frac{p({\cal H}_1)}{p({\cal H}_0)}}_{\hspace*{3mm}\text{Prior odds ratio}}\hspace*{-3mm},
\end{equation}
the detection rule in \eqref{eq_posterioroddstest} can be written as 
\begin{equation}
	\label{eq_likelihoodratiotest}
	L(\mathbf{z}_n)\overset{\Delta}{=}\frac{p(\mathbf{z}_n|{\cal H}_1)}{p(\mathbf{z}_n|{\cal H}_0)}>\gamma
\end{equation}
where the detection threshold is
\begin{align}
\begin{split}
\label{E:gamma}
	\gamma&\overset{_\Delta}{=}\frac{1-p({\cal H}_1)}{p({\cal H}_1)}\cdot\eta\\&=\frac{1-p({\cal H}_1)}{p({\cal H}_1)}\cdot\frac{\lambda_{10}-\lambda_{00}}{\lambda_{01}-\lambda_{11}}.
	\end{split}
\end{align}
Thus, performing a zero-velocity detection using the posterior odds ratio in \eqref{eq_posterioroddstest} is equivalent to performing a likelihood ratio test with a threshold dependent on the hypothesis prior $p({\cal H}_1)$ and the loss factor $\eta$.

As shown in \cite{Skog2010}, many commonly applied zero-velocity detectors can, given different assumptions about the prior knowledge of the sensor signals, be derived as (generalized) likelihood ratio tests of the same form as \eqref{eq_likelihoodratiotest}. These include the acceleration-moving variance detector, the acceleration-magnitude detector, the angular rate energy detector, and the stance hypothesis optimal detection (SHOE) detector. Hence, the threshold in these detectors can be interpreted using (\ref{E:gamma}). What is more, the relationship in (\ref{E:gamma}) provides a theoretically sound way to design adaptive thresholds for these detectors. Next, we discuss how this can be done by letting the hypothesis prior $p({\cal H}_1)$ and the loss factor $\eta$ be time-dependent. 

\subsection{Modeling the Loss Factor}

First, we set the cost of deciding on the correct hypothesis to zero, i.e., $\lambda_{00}=0$ and $\lambda_{11}=0$. The cost of a false alarm (incorrectly deciding that the IMU is stationary) is generally high due to the damaging effect of imposing an erroneous zero-velocity update on the navigation solution \cite{Nilsson2012}. However, the problem of erroneous zero-velocity updates cannot easily be shown to be dependent on other factors, and hence, $\lambda_{10}$ is assumed to be a constant. Moreover, since $\eta=\lambda_{10}/\lambda_{01}$, all time-dependence can without loss of generality be modeled in $\lambda_{01}$. The cost of a missed detection (incorrectly deciding that the IMU is moving) varies more distinctly with time. Immediately after a zero-velocity update, the uncertainty of the observable elements in the navigation state is low \cite{Nilsson2013}, and an additional zero-velocity update will only lead to a marginal improvement in estimation accuracy\footnote{In-fact, as described in \cite{Nilsson2012}, zero-velocity updates themselves induce systematic errors into the system, and excessive use of them may deteriorate the system performance.}. However, the longer the time that has passed since the last zero-velocity update, the more important it is to break the error growth of the inertial navigation. Therefore, $\lambda_{01}$ should increase with the time that has passed since the last detected zero-velocity instance. Further, since the likelihood ratio typically spans many orders of magnitude, a logarithmic or polynomial growth will seem slow and be hard to tune. Thus, in this paper, we will consider $\lambda_{01}$ to have exponential growth. In summary, this means that the loss factor will be modeled to have an exponential decay according to 
\begin{equation}
	\label{eq_lossfactor}
	\eta=\alpha e^{-\theta \Delta t_k}
\end{equation}
where $\alpha$ and $\theta$ are design parameters and $\Delta t_k$ is the time since the last zero-velocity instance. If needed to avoid false alarms, it is possible to set a lower limit on $\eta$, so that $\eta=\max(\alpha e^{-\theta \Delta t_k},\ell)$, where $\ell$ is some design parameter.

\subsection{Modeling the Hypothesis Prior}
\label{subsection_hypothesis_prior}

The hypothesis prior will be dependent on what information is available in each specific scenario. 
In a zero-velocity-aided aided inertial navigation system implemented using a Bayesian filter, the system calculates a navigation solution in terms of a statistical distribution. This statistical distribution can be used to calculate a prior for the detector. Hence, consider the case when an extended Kalman filter is used, which at sampling instant $k$ provides the velocity estimate $\hat{\mathbf{v}}_k$ and velocity error covariance $\mathbf{S}_k$. A measure of how close the system is to have zero velocity, weighted by the uncertainty of the velocity estimate, is then given by $\xi_k=\hat{\mathbf{v}}_k^\top\mathbf{S}_k^{-1}\hat{\mathbf{v}}_k$. Following the ideas of logistic regression, this measure can be mapped to the probability of the system being stationary via the logistic function. Thus, the prior is then set as
\begin{equation}
	\label{eq_ph1}
	p({\cal H}_1)=\frac{1}{1+e^{\beta_1\xi_k+\beta_2}}
\end{equation}
where $\beta_1$ and $\beta_2$ are design parameters. 

For $\xi_k$ to be a reliable measure of how close the system is to have zero velocity, the covariance $\mathbf{S}_k$ must reflect the true uncertainty of the velocity estimate $\hat{\mathbf{v}}_k$. If this cannot be guaranteed, the uninformative prior $p({\cal H}_1)=1/2$ may be a better choice. 


\subsection{Parameter Selection}
\label{subsection_parameter_selection}

Given the cost factor in \eqref{eq_lossfactor} and the prior in \eqref{eq_ph1}, the logarithm of the detection threshold becomes
\begin{equation}
	\label{eg_log_gamma}
    \log \gamma_k=c_1+c_2\Delta t_k+c_3\xi_k
\end{equation}
where $c_1=\beta_2+\log \alpha$, $c_2=-\theta$, and $c_3=\beta_1$. With an uninformative prior, \eqref{eg_log_gamma} still holds but with $c_1=\log\alpha$, $c_2=-\theta$, and $c_3=0$. The parameters $c_1$, $c_2$, and $c_3$ may be estimated using ground truth data. However, this requires training data. Therefore, in what follows, a semi-heuristic parameter selection method is presented.

\begin{figure}[t]
	\vspace*{0mm}
	\hspace*{-2mm}
	\includegraphics{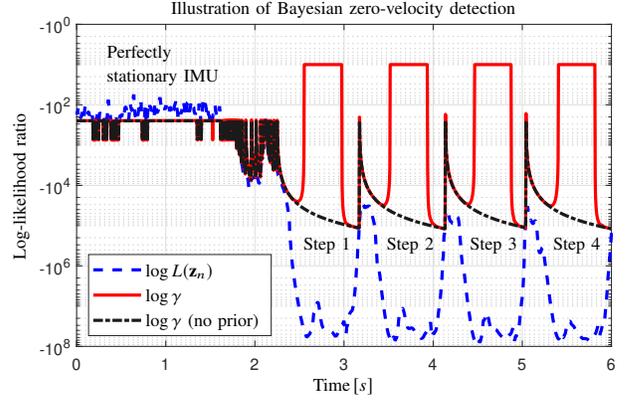}
	\vspace*{-2mm}
	\caption{Illustration of Bayesian zero-velocity detector using the likelihood ratio of the SHOE detector (see \cite{Skog2010}) and an adaptive threshold.  \label{fig:parameterselection}}
	\vspace*{0mm}
\end{figure}

When the IMU is perfectly stationary the detector should with high probability decide on hypothesis $\mathcal{H}_1$. This can be achieved by enforcing the condition $p(\log L(\mathbf{z}^\ast_n)<c_1)=\epsilon$, where $\epsilon$ is small (e.g., $\epsilon=0.05$) and $\mathbf{z}^\ast_n$ is the IMU output when the IMU is perfectly stationary. Here, we have used that $\xi_k$ tends to be small at stationarity. For illustration, examine the first second in Fig. \ref{fig:parameterselection} where frequent zero-velocity instances are detected for a stationary IMU. 

Similarly, it is reasonable to expect a zero-velocity update after completing a step. In mathematical terms, this gives us 
$p(\log L(\mathbf{z}^\diamond_n)<c_1+c_2\Delta\tau)=\epsilon$, where $\Delta\tau$ is the approximate time length of a step and $\mathbf{z}^\diamond_n$ is the output from the IMU during the midstance of normal gait. In this way, most steps during normal walking will result in detected zero-velocity instances. Fig. \ref{fig:parameterselection} illustrates how the threshold and likelihood ratio intersect at the end of each step.

With an uninformative prior, one may select the parameters $c_1$ and $c_2$ by using the two conditions described above and set $c_3=0$. However, with the prior in \eqref{eq_ph1} we need a third condition to perform the parameter selection. To this end, note that the detector should decide on hypothesis ${\cal H}_0$ during the swing phase. 
Therefore, we impose the condition $p(\log L(\mathbf{z}^\star_n)<c_1+c_2\Delta\tau/2+c_3\xi_k^\star)=1-\epsilon$ where $\mathbf{z}^\star_n$ and $\xi_k^\star$ are typical values during the swing phase of $\mathbf{z}_n$ and $\xi_k$, respectively. 
The effect of using the prior in \eqref{eq_ph1} can be seen by comparing the dashed black line and the solid red line in Fig. \ref{fig:parameterselection}.

Finally, note that although the proposed detection framework includes design parameters, it will not be as sensitive to parameter tuning as the conventional detector using a fixed threshold. Firstly, as outlined above, there are thumb rules for tuning the Bayesian detector. Secondly, regardless of how the Bayesian detector is tuned, the risk of missing zero-velocity instances when the user is running (which is the consequence of choosing a too large fixed threshold) is mitigated by the fact that the threshold diminishes with the time since the last detected zero-velocity instance.

\section{Experiments}

The performance of the proposed detector was benchmarked against a detector with a fixed threshold using data from \cite{Skog2010a}. The data set was collected with a MicroStrain 3DM-GX2 IMU and consists of 40 data recordings; half with normal gait ($5\,[km/h]$) and the other half with fast gait ($7\,[km/h]$). In each recording, the user walked one lap along a closed-loop trajectory with an approximate length of $84\,[m]$. The zero-velocity detection was performed using the likelihood ratio of the SHOE detector (see \cite{Skog2010}) with a window length of $20\,[ms]$ (5 samples). The parameters $c_1$, $c_2$, and $c_3$ were quickly tuned based on the procedure outlined in Section \ref{subsection_parameter_selection}, and remained fixed on all data. The two hypothesis priors suggested in Section \ref{subsection_hypothesis_prior} resulted in the same navigation performance. However, for gait with longer periods between the zero velocity events, the informative prior suggested in Section \ref{subsection_hypothesis_prior} may prevent false zero-velocity detections otherwise caused by too aggressive growth of the cost of a missed detection. 

\begin{figure}[t]
	\hspace*{-1.5mm}
	\vspace*{0mm}
	{
		\includegraphics{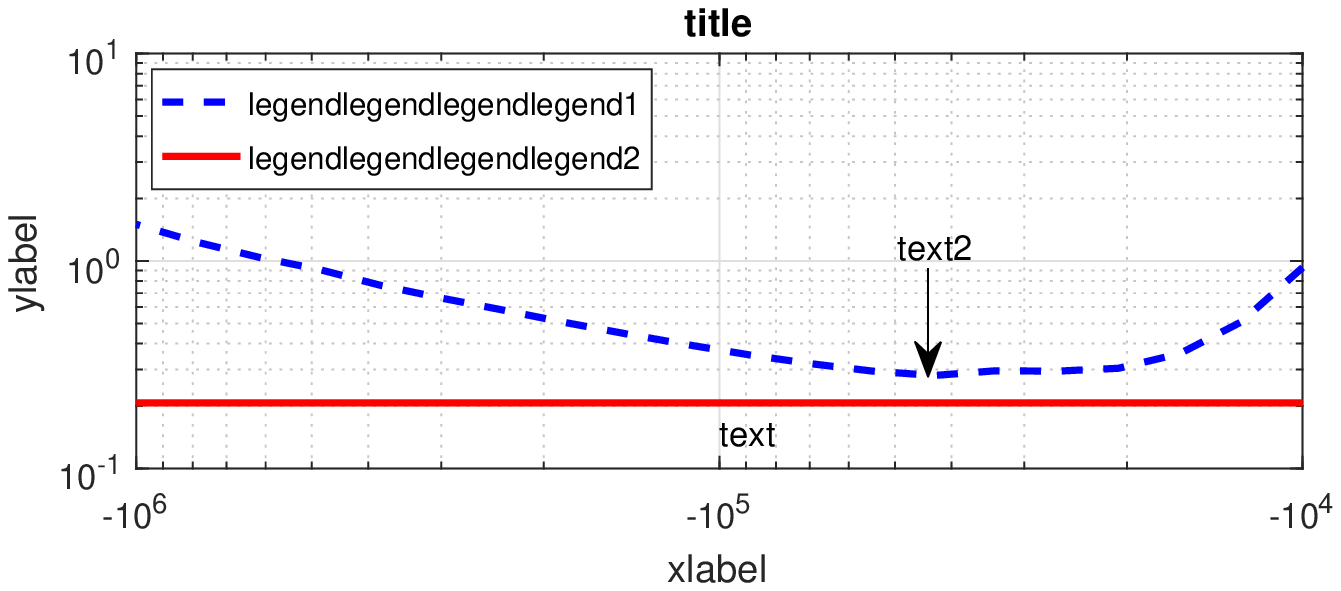}
		\vspace*{2mm}
	}
	\hspace*{-1.4mm}
	{
		\vspace*{0mm}
		\includegraphics{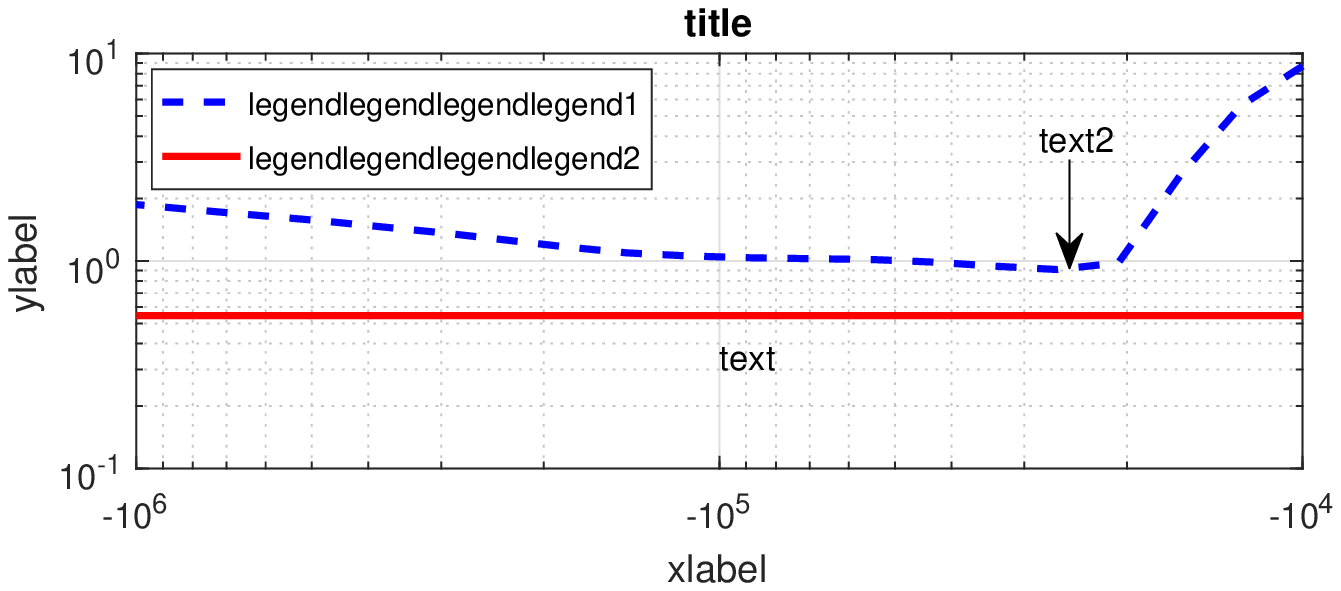}
		\vspace*{2mm}
	}
	\hspace*{-1.4mm}
	{
		\vspace*{-1.3mm}
		\includegraphics{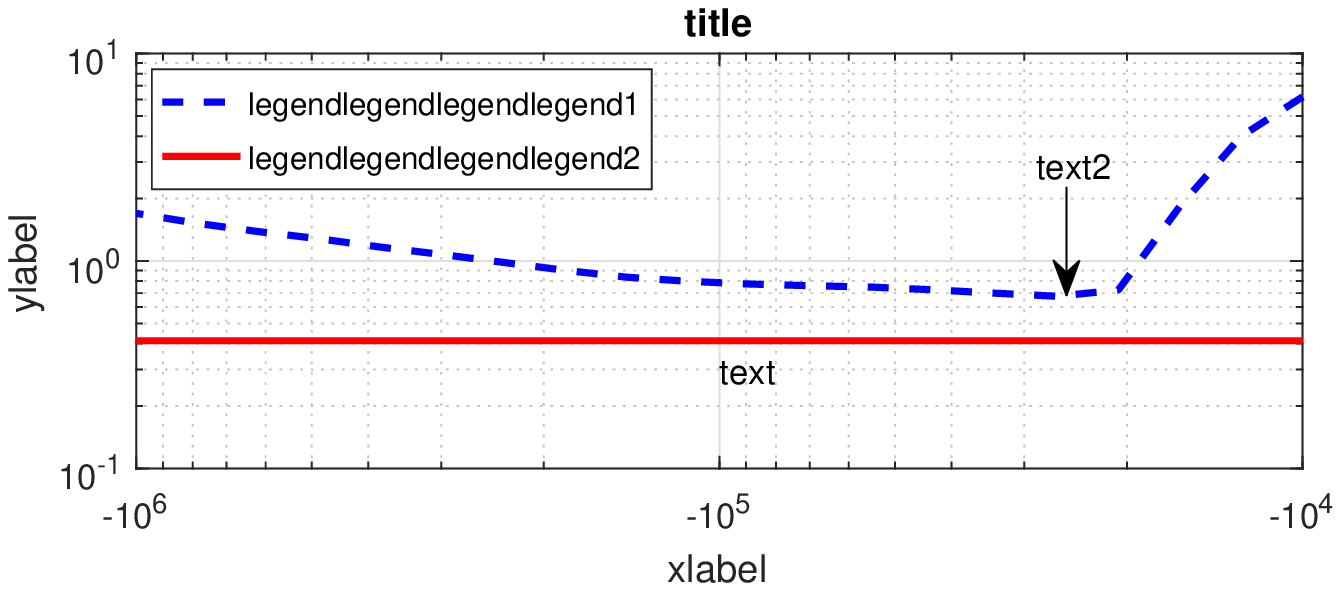}
		\vspace*{0mm}
	}
	\vspace*{-2mm}
	\caption{Position error of zero-velocity-aided inertial navigation using a fixed and an adaptive threshold.\label{fig:experiment}}
	\vspace*{0mm}
	\label{fig1}
\end{figure}

Fig.~\ref{fig:experiment} displays root-mean-square errors (RMSEs) computed from the final position estimates in each trajectory for both the Bayesian detector and the corresponding detector with a fixed threshold. The adaptive threshold is seen to provide a lower RMSE than the best fixed threshold. This holds true not only when performing the evaluation on all data, but also individually on the data with normal gait and the data with fast gait. Thus, these results indicate that the Bayesian detector proposed in this paper is preferable to previously investigated methods for adaptive thresholding that choose a separate threshold for each speed or motion mode, see e.g., \cite{Tian2016,Park2016,Zhang2017,Ma2017,Wagstaff2017}. To illustrate that the difference in performance becomes rather substantial over longer time periods, we also computed the position error resulting from concatenating all available data recordings (corresponding to walking a distance of about $3.4\,[km]$). In this case, the adaptive threshold and the best fixed threshold give position errors with magnitudes of $8.16\,[m]$ and $20.35\,[m]$, respectively.

The performance improvement may be explained in several ways. On the one hand, the adaptive threshold prevents the navigation system from running too long without a zero-velocity update, and thereby prevents the estimation errors from growing too large before the next zero-velocity update. On the other hand, the adaptive threshold prevents the navigation system from using an excessive number of zero-velocity updates, and thereby reduces the negative impact that modeling errors in the pseudo velocity measurements have on the navigation solution.

\section{Conclusion and discussion}

This paper has developed a Bayesian zero-velocity detector using the posterior odds ratio. Experiments using data collected at varying gait speeds indicated that the proposed detector provides a significantly lower positioning error than state-of-the-art detectors where an adaptive threshold is selected based on the motion mode of the user. Further, the proposed method does not rely on learning the relationship between the user motion and the optimal threshold, and therefore avoids the time-consuming data collection that has been necessary in previous methods for adaptive thresholding. 

One possible direction for future work is to evaluate the Bayesian detector for a broader set of motion classes and gait conditions. Although our results indicate that the Bayesian detector alleviates the need for adaptive parameter tuning, the Bayesian detector could, in the same way as the conventional detector with a fixed threshold, still be used together with motion classifiers that enable adaptive parameter tuning.

\section*{Acknowledgment}
\addcontentsline{toc}{section}{Acknowledgment}
\scriptsize
This research has been financially supported by the National Institute of Standards and Technology (NIST) via the grant \emph{Pervasive, Accurate, and Reliable Location-based Services for Emergency Responders} (Federal Grant: 70NANB17H185).

\normalsize

%
%

%
%
%
%
\bibliographystyle{IEEEtran}

\bibliography{IEEEabrv,refs}

\end{document}